%% file: main.tex
\documentclass[10pt,journal,twocolumn,twoside]{IEEEtran}
\usepackage[utf8]{inputenc}
\usepackage[T1]{fontenc}

\usepackage{booktabs} 

\usepackage[caption=false,font=footnotesize]{subfig}
\usepackage{floatrow}
\usepackage{fancyhdr}
\usepackage{mdwlist}
\usepackage{graphicx}
\usepackage{cite}
\usepackage{color}
\usepackage{multirow}
\usepackage{amsmath}        

\graphicspath{{incl/}}

\makeatletter
\def\blfootnote{\xdef\@thefnmark{}\@footnotetext}
\makeatother

\definecolor{gray}{gray}{0.9}

\newcommand{\blue}[1]{\textcolor{blue}{#1}}
\renewcommand{\blue}[1]{#1}

\newcommand{\drop}[1]{}

\begin{document}

\title{%
Toward Physically Unclonable Functions from
Plasmonics-Enhanced Silicon Disc Resonators%
}

\makeatletter
\let\thetitle\@title
\makeatother

\author{Johann~Knechtel,~\IEEEmembership{Member,~IEEE},
	Jacek~Gosciniak,
	Alabi~Bojesomo,
	Satwik~Patnaik,~\IEEEmembership{Student Member,~IEEE},
        Ozgur~Sinanoglu,~\IEEEmembership{Senior Member,~IEEE},
        and Mahmoud~Rasras,~\IEEEmembership{Senior Member,~IEEE},
\thanks{J.\ Knechtel and J.\ Gosciniak contributed equally.
	J.\ Knechtel, J.\ Gosciniak, A.\ Bojesomo, O.\ Sinanoglu, and M.\ Rasras are 
with the Division of Engineering, New York University Abu Dhabi,
Saadiyat Island, 129188, UAE (e-mail: johann@nyu.edu; jg5648@nyu.edu; asb701@nyu.edu; ozgursin@nyu.edu; mrasras@nyu.edu).
S.\ Patnaik is with the Department
of Electrical and Computer Engineering, Tandon School of Engineering, New York University, Brooklyn, NY, 11201, USA (e-mail: sp4012@nyu.edu).}
}

\markboth{Journal of Lightwave Technology}{Knechtel \MakeLowercase{\textit{et al.}}: \thetitle}

\IEEEtitleabstractindextext{

\begin{abstract}
The omnipresent digitalization trend
has enabled a number of related malicious activities,
ranging from data theft to disruption of businesses, counterfeiting of devices, and identity fraud, among others.
Hence, it is essential to implement security schemes and 
to ensure the reliability and trustworthiness of electronic circuits.
Toward this end, the concept of \textit{physically unclonable functions (PUFs)}
has been established at the beginning of the 21st century.
However, most PUFs
have eventually, at least partially, fallen short of their promises, which are
unpredictability, unclonability, uniqueness, reproducibility, and tamper resilience.
That is because most PUFs directly utilize the
underlying microelectronics, but that
intrinsic randomness can be limited and may thus be predicted, especially by machine learning.
Optical PUFs, in contrast, are still considered as promising---they
can derive strong, hard-to-predict randomness independently from microelectronics, by using some kind of ``optical token.''
Here we propose a novel concept for plasmonics-enhanced optical PUFs, or \textit{peo-PUFs} in short.
For the first time, we leverage two highly nonlinear phenomena in conjunction by construction:
(i)~light propagation in a silicon disk resonator, and 
(ii)~surface plasmons arising from nanoparticles arranged randomly on top of the resonator.
We elaborate on the physical phenomena, provide simulation results,
and conduct a security analysis of peo-PUFs
for secure key generation and authentication.
This study highlights the good potential of peo-PUFs, and our future work is to focus on fabrication and
characterization of such PUFs.
\end{abstract}

\begin{IEEEkeywords}
Hardware Security, Physically Unclonable Function, Plasmonics, Optical Waveguide, Silicon Disc Resonator
\end{IEEEkeywords}
}

\maketitle

\IEEEdisplaynontitleabstractindextext

\renewcommand{\headrulewidth}{0.0pt}
\thispagestyle{fancy}
\lhead{\vspace{-1cm}}
\rhead{\vspace{-1cm}}
\chead{
\copyright~2019 IEEE.
This is the author's version of the work.
It is posted here for your personal use.
Not for redistribution.
The definitive Version of Record is published in
IEEE/OSA J.\ Lightwave Technology (JLT), 2019,
	http://doi.org/10.1109/JLT.2019.2920949
}
\cfoot{}

\section{Introduction}
\label{sec:introduction}

\IEEEPARstart{F}{or}
many decades now, authentication and other security schemes have leveraged the randomized manifestations of selected physical,
    biological, or other phenomena.
For example, biometric identification is based on the unique patterns of fingerprints, retinas, voices, or even
walking pace and motion.
For electronic circuits, the notion of \textit{physically unclonable functions (PUFs)} has been established at the beginning of the 21st
century~\cite{pappu02,maes10,herder14,chang17_PUF}.
When applied some input stimulus,
a PUF should provide a randomized, fully de-correlated output
response. This response should be reproducible for the very same PUF, even under varying environmental conditions, but it should differ 
across different PUF instances, even for the same PUF design.
PUFs are used for (i)~challenge-response-based security schemes, which require capabilities for processing a large number of inputs (using
		so-called ``strong PUFs'')
or (ii)~for ``fingerprinting'' or simple key generation schemes, which require capabilities for processing only one or few fixed
inputs (using so-called ``weak PUFs'')~\cite{herder14,maes10,chang17_PUF}.

In short, the desired properties for a PUF are uniqueness, unclonability, unpredictability, reproducibility, and tamper resilience, while
possible applications are device authentication, secure generation of keys, and device-entangled cryptography.

The core principle for electronic PUFs is to leverage the process variations inherent to microelectronic fabrication
and to boost these variations purposefully using some dedicated circuitry. 
Prominent types of electronic PUFs are ring oscillators, arbiters, bistable rings, and
memory-based PUFs~\cite{maes10,herder14,ganji17_thesis,chang17_PUF}.
Such PUFs are relatively simple to implement and integrate, even for advanced processing nodes.
However,
	it is also important to note that
	all these PUFs rely on the intrinsic randomness of the underlying microelectronics. This randomness
can be limited and may be eventually predicted/cloned.
In fact, various attacks have been demonstrated,
with machine learning emerging as the most powerful approach~\cite{chang17_PUF,ruehrmair13,liu17,ganji17_thesis}.

Another interesting option are \textit{optical PUFs}~\cite{pappu02,ruehrmair13_IACR,tuyls07,maes10,grubel17}.
Here the idea is to
manufacture an ``optical token'' which, in addition to structural variations inherently present in selected optical media, 
may contain randomly included materials (e.g., microscopic particles).
Besides such a token, optical PUFs require further components, for generating the optical input and processing the output.
The fundamental phenomena underlying an optical PUF are scattering, reflection, coupling, and absorption of light within the optical token.
Depending on the materials used for the token and the inclusions, as well as the design of the token itself, these phenomena can be highly
chaotic by nature~\cite{grubel17,kauranen12}.
Hence, optical PUFs are in principle more powerful than other types of PUFs.
Still, prior art on optical PUFs has some practical limitations, e.g., the use of linear media, external and exposed optical tokens,
	the need for complex and sensitive setups,
	or the need to customize manufacturing steps for different PUF tokens.\footnote{See also
	Section~\ref{sec:prior_art} for a more detailed review of prior art.}

Aside from optical PUFs, we note that particles of different type and size have been used for some time for ``tagging'' and optical
authentication of
goods~\cite{ruehrmair12, smith17_plasmonics}. In fact, the first well-known approach for
secure authentication of goods was particle-based tagging of nuclear weapons during the cold war~\cite{ruehrmair12}.
Different materials can be leveraged for particle-based tagging, e.g., quantum dots, fluorescent particles, or metallic nanoparticles (NPs).
The latter are particularly interesting, as they effectively boost the light-matter interaction.
Indeed, the concept of \textit{plasmonic NPs} has gained
significant traction recently for security
schemes~\cite{smith17_plasmonics,zheng16_plasmonics,smith16_plasmonics,park16_plasmonics,cui14_plasmonics}.
As with most optical PUFs, however, current schemes require external components;
integrated schemes have not been proposed yet.

\begin{figure}[tb]
\centering
	\includegraphics[width=\columnwidth]{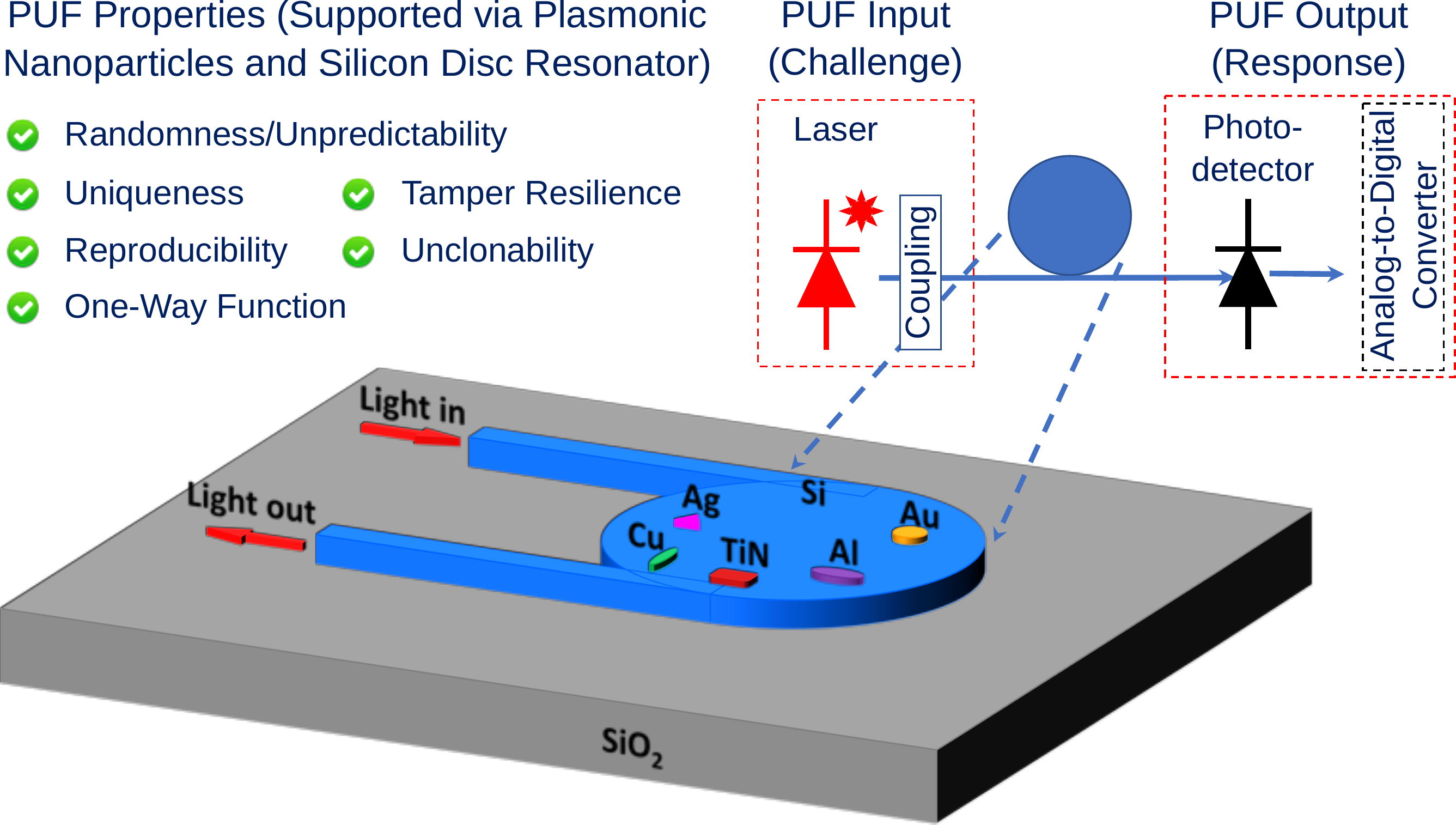}
\caption{
		Our concept for peo-PUFs.  At its heart is an device-integrated, micrometer-sized silicon disk resonator with plasmonic
			NPs of different shapes, sizes, and metals, randomly arranged on top of the resonator.  Considering recent
			advances for optoelectronics, the peripheral components may be at least partially integrated onto the electronic
			device as well.
\label{fig:scheme}
}
\end{figure}

Given that plasmonic NPs can induce highly nonlinear behavior~\cite{kauranen12,park16_plasmonics}---which is extremely valuable
when designing
a PUF~\cite{ganji17_thesis}---it is surprising that \textit{none} of the prior works did consider such NPs for an advanced optical PUF.
This paper can be summarized as follows:
\begin{itemize}
\item For the first time concerning PUFs,
we propose to entangle two promising, highly nonlinear physical phenomena by construction:
light propagation in a silicon disk resonator, and 
surface plasmons arising from NPs arranged randomly on top of the resonator.
We name this concept as \textit{peo-PUFs}, short for plasmonics-enhanced optical PUFs (physically unclonable functions).
The concept is illustrated in Fig.~\ref{fig:scheme}.
\item We discuss and study the underlying physical phenomena, the latter by
means of \textit{Lumerical FDTD} and
\textit{COMSOL} simulations.
Using the same tools, we obtain different data sets for various peo-PUFs under different conditions.
\item Based on these data sets, we evaluate the randomness, uniqueness, and reproducibility of peo-PUFs,
confirming their applicability for secure key generation in principle.
We also propose a simple authentication scheme, based on the secure keys and some related helper data.
\item The concept can be directly integrated into any silicon device,
thus it can provide better resilience than prior external PUFs.
The manufacturing of silicon disc resonators is a well-known process, and
plasmonic NPs can be easily deposited on top of a resonator in a random fashion,
e.g., by sputtering.
We believe that these two aspects are essential for successful adoption of peo-PUFs.
\end{itemize}

\section{Prior Art for Optical PUFs}
\label{sec:prior_art}

We like to point out that the very first PUF proposal was actually an optical PUF;
in 2002, Pappu \textit{et al.\ }\cite{pappu02} made an optical token 
from transparent epoxy with randomly inserted, micrometer-sized glass spheres.
That token was illuminated by an external laser, whereupon
the resulting speckle pattern was visually recorded, filtered, and digitized.
In 2013, R\"uhrmair \textit{et al.\ }\cite{ruehrmair13_IACR} first replicated
and confirmed the findings by Pappu \textit{et al.\ }and then
prototyped an integrated optical PUF based on the same working principle.
Tuyls \textit{et al.\ }\cite{tuyls07} discussed integrated optical PUFs in 2017, albeit only in theory, without any experimental evaluation.
Also in 2017, Grubel \textit{et al.\ }\cite{grubel17} demonstrated an resonator-based PUF with pseudo-randomized
structures.  To the best of our knowledge, their work was the first to demonstrate a nonlinear optical PUF.

While these works are promising,
there are also some notable limitations. As for the early external optical PUFs~\cite{pappu02,ruehrmair13_IACR},
their setups are relatively complex. Thus, these PUFs are
not only sensitive to environmental
parameters like variations of temperature and supply voltages, which is the case for any type of
PUF, but also to mechanical vibrations, laser alignment, etc.
Besides, exposing the token can result in wear and tear;
an external PUF may become irreproducible after some time.
Even more concerning, an external PUF can arguably never be trusted completely---an attacker can
take hold of the token and, subsequently, (a)~re-use it for authenticating of counterfeit chips, or
(b)~explore its challenge-response behavior for modeling attacks.
As for the early integrated optical PUFs~\cite{ruehrmair13_IACR,tuyls07}, their shortcoming is the limitation to linear materials which can
be modeled/cloned~\cite{ruehrmair13_IACR}.
The recently proposed PUF~\cite{grubel17}, while certainly an advancement over the other prior art, has still limitations.
For one, it requires sophisticated, external optoelectronic components, e.g.,
pulse pattern generators and programmable spectral filters.
Moreover, this PUF relies on pseudo-randomized structures within the resonator which requires customizing
the manufacturing steps for different PUFs, which seems impractical.

\section{Discussion and Study of the Phenomena}
\label{sec:physics}

\subsection{Light Propagation in Silicon Resonators}
\label{sec:light_silicon}

It is well-known that silicon is transparent to infrared light and has a very high refractive index.
The accordingly strong photonic confinement allows for the
design of micrometer-scale, yet efficient, silicon-made optical waveguides~\cite{li18_photonics}.
Among others, prominent applications are optoelectronic filters based on
\textit{silicon disc resonators (SDRs)}, i.e., circular cavities
which enable strong light interaction.
That is, light entering an SDR builds up in intensity over multiple round-trips within the cavity,
where only particular wavelengths, depending on the SDR design, will be in resonance.

The light propagation within SDRs in particular and silicon devices in general are subject to various nonlinear
effects. These include the \textit{Kerr effect}, \textit{Raman scattering}, \textit{self-modulation},
\textit{two-photon absorption}, et cetera\blue{~\cite{li18_photonics, kauranen12, grubel17, wang2013multi}}.
The nonlinearity in silicon is fundamental for light-to-light interaction, which itself is essential for techniques such as wavelength
conversion.
Depending on the SDR design, other chaotic effects like \textit{dynamic billiards} can also play a role~\cite{grubel17}.

\subsection{Plasmonics}
\label{sec:plasmonics}

Because of the outstanding capability for sub-wavelength confinement of electromagnetic energy, plasmonics has become a driving force for
progress in the area of nanophotonics\blue{~\cite{gosciniak2019plasmonic, gosciniak2019high, kumar2013dielectric}}.
Metallic nanostructures are at the heart of plasmonics---the phenomenon of plasmonics originates from
strong coupling of photon energy with free electrons in a metal.
This strong coupling supports a wave of charge-density fluctuations along the surface of the metal, thereby creating a sub-wavelength
oscillating mode called a \textit{surface plasmon}\blue{~\cite{bozhevolnyi09_plasmonics,chen2011,kauranen12}}.
More specifically, for one, there is the electromagnetic energy transport for the propagation of \textit{surface plasmon polaritons (SPPs)}
sustained at the planar metal/dielectric interface;
for another, there is the \textit{localized surface plasmon resonance (LSPR)}, a non-propagating excitation of the metal's conduction electrons
coupled to the electromagnetic field.
Note that the LSPR in particular is under intensive research for many years
now\blue{~\cite{wang06_plasmonics,schuller10_plasmonics,park16_plasmonics,kauranen12,gosciniak17, scholl2012quantum}};
that interest is also because a large
variety of metallic nanostructures are commercially available, which all give rise to
unique properties and applications.

\begin{figure}[tb]
	\centering
\sidesubfloat[]{
	\includegraphics[height=39mm, trim = {1mm 0mm 6mm 1mm}, clip=true]{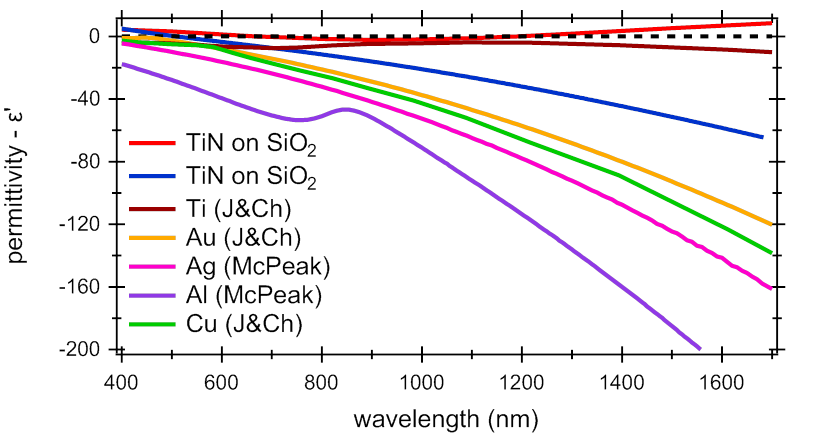}
}
\\[2mm]
\sidesubfloat[]{
	\hspace{2mm}\includegraphics[height=38mm, trim = {1mm 0mm 6mm 1mm}, clip=true]{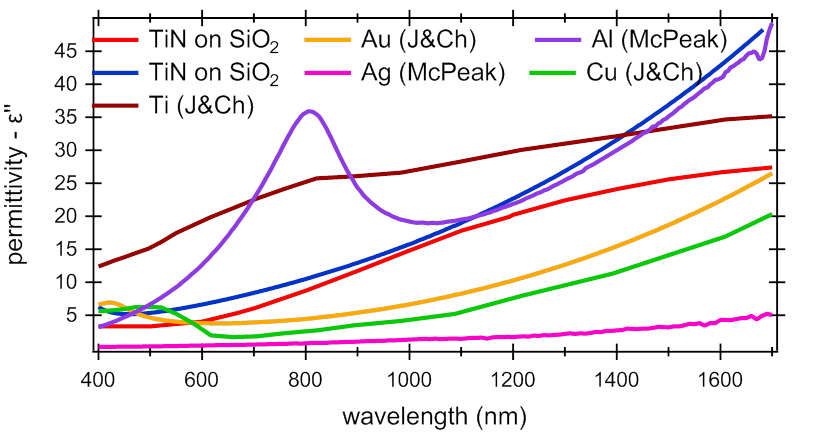}
}
\caption{Permittivities of metals used in plasmonics, with their (a)~real and (b)~imaginary parts.
	The two \textit{TiN on SiO2} data sets are for different depositioning conditions.
		For \textit{J\&Ch} and \textit{McPeak} data, see~\cite{permittivities}.
\label{fig:permittivity}
}
\end{figure}

\begin{figure}[tb]
\centering
\includegraphics[width=\columnwidth]{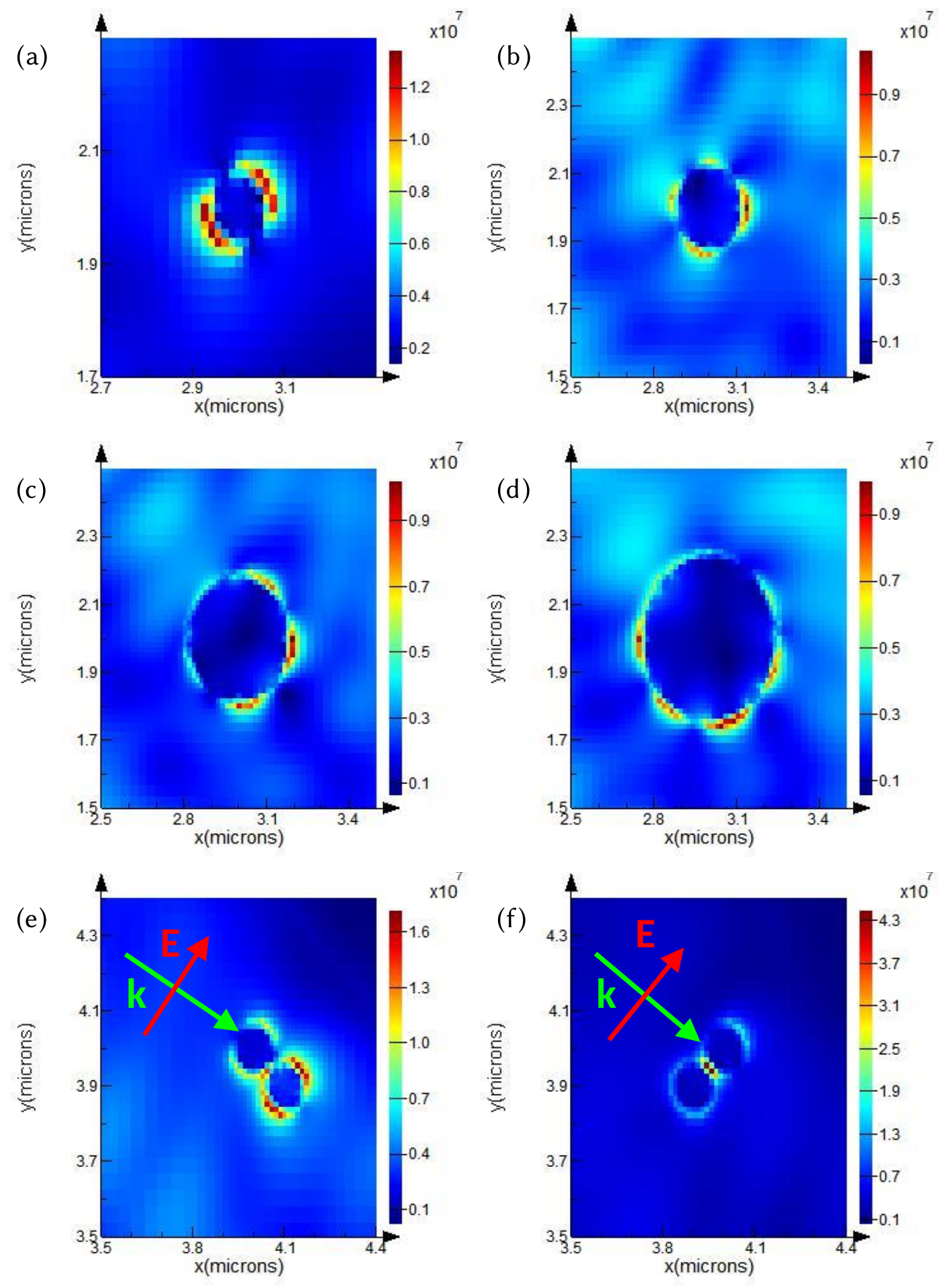}
\caption{Simulations for plasmonic field enhancement.
	(a)--(d): Excitations by single gold NPs of different sizes.
	(e), (f): Excitations by two gold NPs (same size) which are (e) not coupled versus 
		  (f) coupled. The latter shows
		    a strong non-resonant gap mode.
			 Vectors \textit{E} are for the electric field and \textit{k} for the light propagation, respectively.
\label{fig:modes}
}
\end{figure}

The LSPR field enhancement is dictated by various factors.
First and foremost, the enhancement depends
on the metal properties~\cite{gosciniak17}.
For example, Fig.~\ref{fig:permittivity} illustrates the
permittivities of commonly considered metals, all exhibiting highly wavelength-dependent properties.
Moreover, the field enhancement also depends on the NP structure and size\blue{~\cite{wang06_plasmonics, schuller10_plasmonics,
	huang2007plasmonic, noguez2007surface}}, the
coupling arrangement or direction\blue{~\cite{gosciniak17}}, and the materials surrounding the NP~\cite{smith16_plasmonics}. See 
Fig.~\ref{fig:modes} for simulations with differently sized and arranged NPs; the simulation setup is described in Section~\ref{sec:setup}.

\begin{figure}[tb]
\centering
\sidesubfloat[]{
	\includegraphics[height=39mm, trim = {0mm 0mm 6mm 1mm}, clip=true]{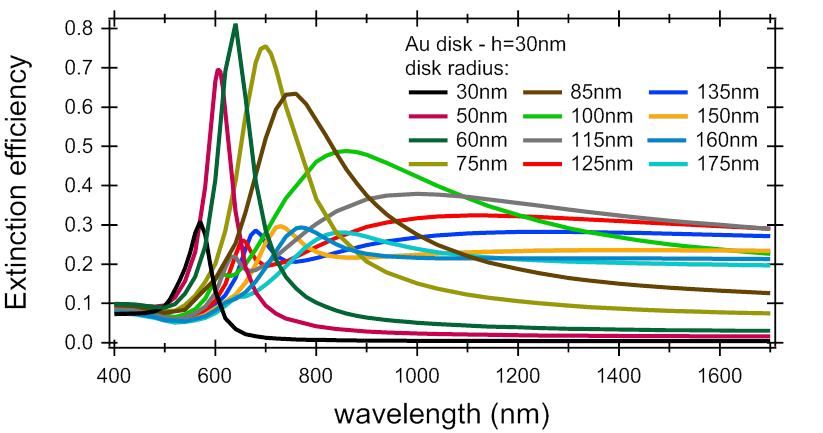}
}
\\[2mm]
\sidesubfloat[]{
	\includegraphics[height=39mm, trim = {0mm 0mm 6mm 1mm}, clip=true]{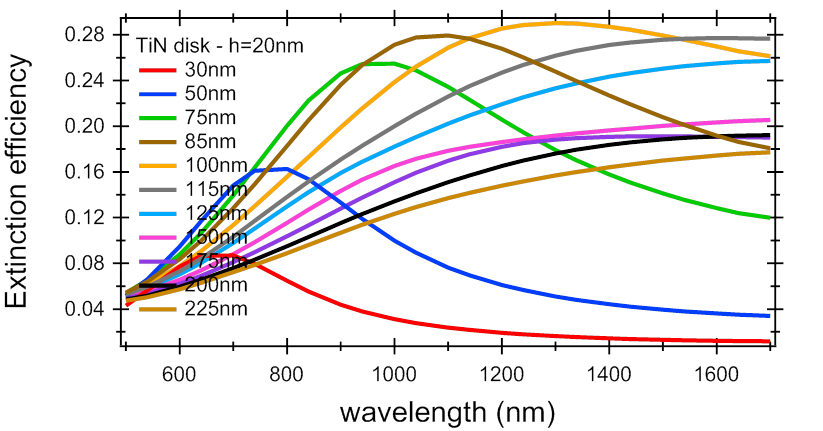}
}
\caption{Extinction efficiencies for (a) gold and (b) titanium nitride NPs of different sizes, but uniform height/thickness (30nm and 20nm,
		respectively). NPs are labelled as ``disk'' in the plots.
\label{fig:extinction}
}
\end{figure}

Besides LSPR field enhancement, when light interacts with a NP, the light can be absorbed and/or scattered.
Those processes arise in resonant conditions, i.e., where the absorption efficiency is the highest.
The \textit{extinction efficiency}, that is the sum of absorption and scattering efficiencies,
can be engineered via the size, shape, and dielectric environment of the metallic NPs
(Fig.~\ref{fig:extinction}\blue{; see below for details}).
For NPs smaller than the wavelength of excitation, note that the efficiency of absorption dominates over
scattering.
For NPs that have one or more dimensions approaching the excitation wavelength, the optical
phase can vary across the structure. Thus, the retardation effect should be accounted for.
Such nanostructures can also be considered as SPP waveguides that propagate back and forth between the metal terminations, thereby creating a
\textit{Fabry-Perot resonator} for SPPs. 

\blue{The extinction efficiencies illustrated in Fig.~\ref{fig:extinction} were obtained via
	\textit{COMSOL FEM} simulations.
The electric field of light is polarized along the side of the NPs, i.e.,
for the transverse electric mode propagating in the SDR.
For such an arrangement, the dipole or higher-order plasmonic modes of the NPs are excited;
for small NPs, only the dipole mode is supported.
As the NP radius increases, the higher-order modes appear, which shift the resonant conditions to longer 
wavelengths (see curve peaks).
At the same time, higher-order modes can also appear at shorter wavelengths (e.g., Fig.~\ref{fig:extinction}(a), for NP radius 135nm).
The extinction efficiency and resonance conditions depend not 
only on the NP size but also the material properties; recall Fig.~\ref{fig:permittivity}.
The shape of the extinction-efficiency curve depends mostly on the resonance conditions of an NP.  For gold NPs
(Fig.~\ref{fig:extinction}(a)), the extinction efficiency is narrow, with the maximum located at the wavelength of 640nm, for an NP with radius
		of 60nm.  In contrast, the extinction  efficiency for titanium-nitride NPs (Fig.~\ref{fig:extinction}(b)) is much broader and
shifted toward longer wavelengths, with the maximum located at the wavelength of 1300nm, for an NP with radius of 100nm.}

Other non-resonant effects
are to be considered as well.
Most important, for two or more NPs that are close to each other,
the longitudinal field component of the SPP can ``jump inside the
gap'' such that
the field is particularly enhanced within the gap (Fig.~\ref{fig:modes}(f)).
The enhancement magnitude depends
on the size of the NPs and the spacing between them:
the smaller the gap, the higher the enhancement.
In case the field component is parallel to or outside of the gap,
each of the NPs excites its ``own'' mode which are not coupled (Fig.~\ref{fig:modes}(e)).

\section{Toward Plasmonics-Enhanced Optical PUFs}
\label{sec:peo-PUF}

\subsection{Concept}

Considering all these effects which we outlined above,
it is intuitive that metallic NPs on top of an SDR will significantly impact its photonic mode
propagation. This, in turn, will help to induce highly nonlinear behavior for our envisioned PUF concept.
As outlined, the disturbances arise primarily through absorption, scattering, and LSPR field enhancement, but with all effects acting at once.
Given that the SDR will be made from silicon, a nonlinear material, the
field enhancement can also locally impact the SDR's refractive index,
thereby further disturbing the mode propagation.
As with plasmonic field enhancement, this interference depends on the size of the NPs, 
their coupling direction, light polarization, material properties, et cetera~\cite{smith16_plasmonics}.

In this work, for the first time, we entangle the two nonlinear phenomena of silicon photonics and plasmonics.
We propose the concept of \textit{plasmonics-enhanced optical PUFs (peo-PUFs)}.
Besides the concept in general, we furthermore
propose to leverage peo-PUFs, for now, as so-called ``weak PUFs'' for secure key generation and authentication.
There are two important aspects to note here.
\begin{enumerate}
\item
``Weak PUFs'' are not necessarily inferior to ``strong'' PUFs~\cite{ruehrmair13}.
On the contrary, powerful machine learning attacks such as~\cite{chang17_PUF,ruehrmair13,liu17,ganji17_thesis,atakhodjaev18} do \textit{not}
apply for weak PUFs, only for strong PUFs.
The main difference between weak and strong PUFs is that the former work on a (few) fixed input(s), or \textit{challenge(s)}, whereas the
latter have to support a large range of inputs/challenges.
\item Peo-PUFs may also be implemented as strong PUFs; the SDR tokens by themselves can readily support a large range of optical inputs.
However, this would necessitate more complex optoelectronics for the PUF devices.\footnote{
	Besides,
   in anticipation of related security concerns,
we would like to stress that Atakhodjaev \textit{et al.\ }\cite{atakhodjaev18} have
recently shown that machine-learning attacks on strong SDR-based PUF can be already challenging,
due to the inherent nonlinearity of silicon SDRs.
As we motivate in this paper, once the additional phenomena of plasmonics becomes intertwined, we can reasonably expected even better
resilience against such attacks.}
\end{enumerate}

It is important to note that, as of now, our work is carried out
at the level of modelling, physical simulations, and analytical security evaluation.
For future work, we will focus on the manufacturing and characterization of peo-PUFs as well as the verification of security promises in
manufactured peo-PUFs.

Next, we outline the design, working principle, and key generation and authentication for peo-PUFs.
The simulation setup is given in Section~\ref{sec:setup}.

\subsection{Optical Token and Optoelectronics}
\label{sec:optoelectronics}
Recall that silicon disc resonators (SDRs for short) are at the heart peo-PUFs,
upon which plasmonic NPs of various shapes, sizes, and metals are 
placed randomly (Fig.~\ref{fig:scheme}).
The manufacturing of silicon waveguides/resonators as well as the depositioning of NPs are well-established processes.
Unlike most prior art for optical PUFs, peo-PUFs can, therefore, take full advantage of commercial manufacturing facilities.
Moreover, the token can be irrevocably integrated on the electronic device---we argue that this is essential for proper 
implementation of security schemes.

We argue that advances for optoelectronics may allow us to strive for monolithic integration sooner than later.
For example, IBM has recently demonstrated an ultra-fast photonic intra-chip communication link~\cite{Xiong16}.
In general, the monolithic integration of optical modulators and photodetectors is considered mature, with research focus shifting
toward the integration of the light source/lasers~\cite{seifried18, Guan18, orcutt12, soref06}.
Besides, hybrid integration of plasmonics and silicon photonics has been demonstrated as well, e.g., by Chen \textit{et al.\ }\cite{chen18_plasmonics}.

As indicated, while peo-PUF tokens can support broad ranges of optical inputs in principle, generating these inputs would require, e.g.,
      spectral encoding, frequency sweeping, or power modulation.  Therefore, the challenge here would be to manage the complexity of the
      related optoelectronics, which may render a monolithic integration of peo-PUFs ultimately more difficult.  Hence, in the remainder, we
      assume a simple setup with a fixed laser input pulse for the peo-PUFs.

\subsection{Working Principle}
Once an optical pulse enters the SDR token, the nonlinear phenomena of silicon photonics and plasmonics
result in a highly complex and chaotic,
yet reproducible, optical output.

\begin{figure}[tb]
\centering
\sidesubfloat[]{
	\includegraphics[width=.43\columnwidth, trim = {16mm 9mm 8mm 1mm}, clip=true]{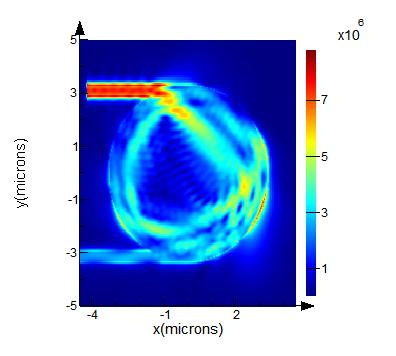}
}\hfill
\sidesubfloat[]{
	\includegraphics[width=.43\columnwidth, trim = {16mm 9mm 8mm 1mm}, clip=true]{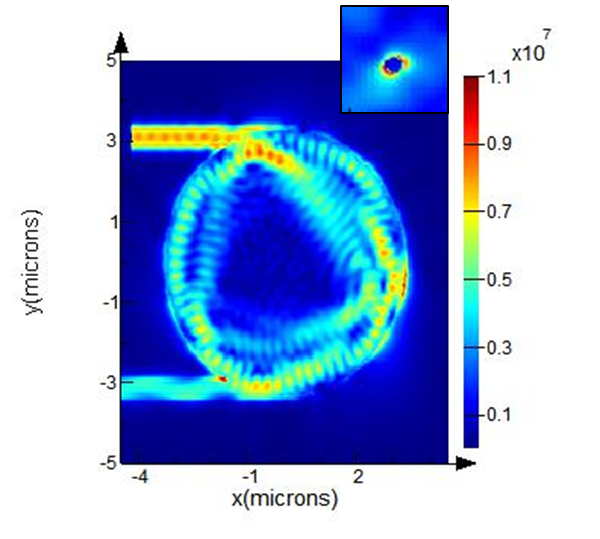}
}
\caption{Electric fields and photonic propagation for an SDR without NPs (a) versus with one gold NP (b).
	\blue{The inset in (b) shows the propagation mode for the NP.}
	Dimensions in $\mu m$.
\label{fig:field}
}
\end{figure}

\begin{figure}[tb]
\centering
\sidesubfloat[]{
	\includegraphics[height=42mm, trim = {9mm 0mm 5mm 1mm}, clip=true]{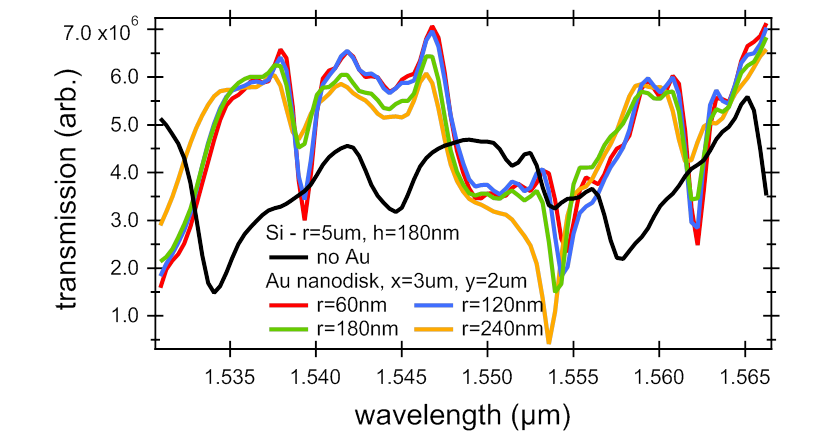}
}
\\[2mm]
\sidesubfloat[]{
	\includegraphics[height=42mm, trim = {9mm 0mm 5mm 1mm}, clip=true]{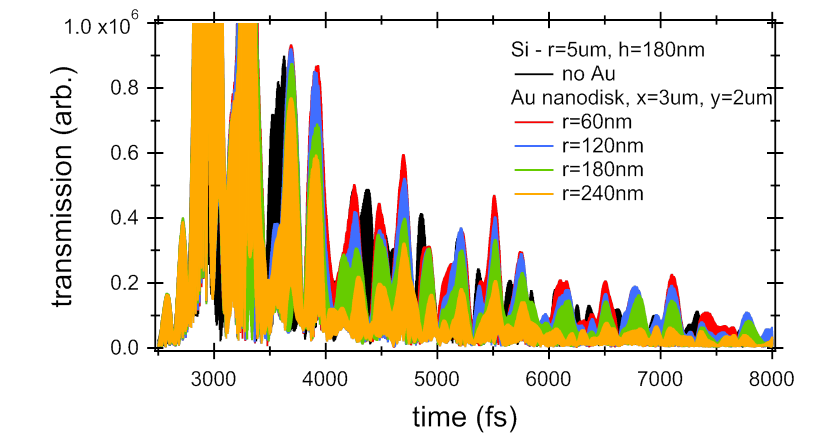}
}
\\[4mm]
\sidesubfloat[]{
	\includegraphics[height=42mm, trim = {9mm 0mm 5mm 1mm}, clip=true]{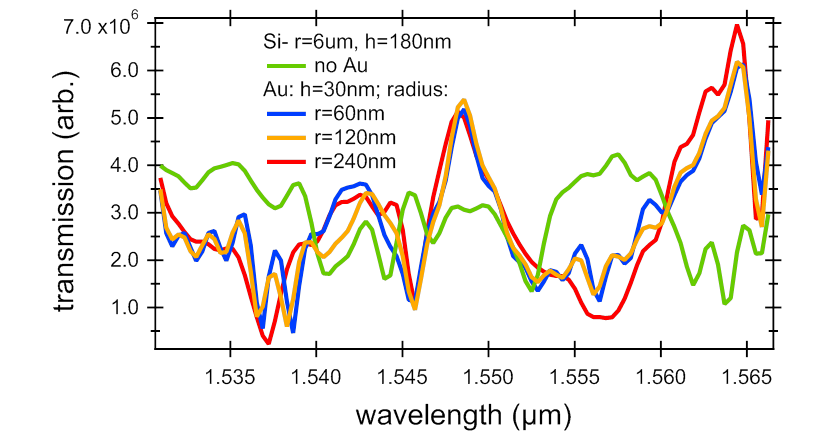}
}
\\[2mm]
\sidesubfloat[]{
	\includegraphics[height=42mm, trim = {9mm 0mm 5mm 1mm}, clip=true]{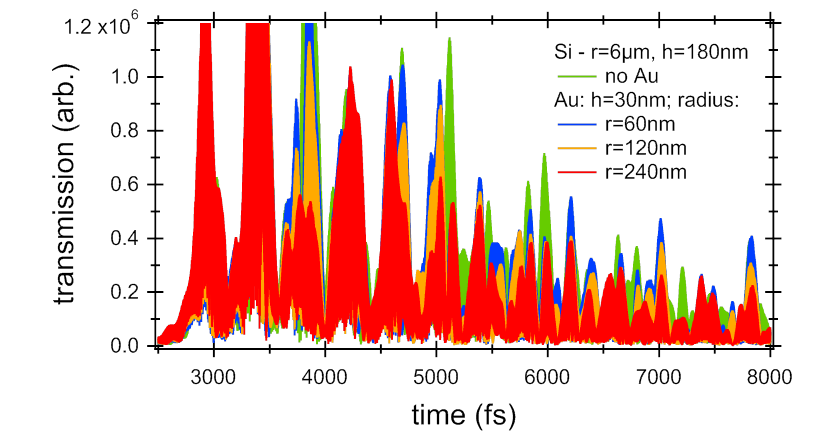}
}
\caption{Transmission plots for different peo-PUF tokens: (a, b) SDR token with radius 6$\mu m$, (c, d) SDR token with radius 5$\mu m$. The SDR's
	height/thickness is 180nm for both setups.
	\blue{Both SDR tokens are independently simulated for three variations of one gold NP, regarding the NP sizes/radius. Without loss of
		generality, the considered NP radii are 60nm, 120nm, and 240nm.}
	For fair comparison, the NPs are always placed at the same location on the SDR.
\label{fig:transmission}
}
\end{figure}

Figure~\ref{fig:field} illustrates the electric field and the photonic propagation mode for an SDR acting as peo-PUF token.
\blue{Parameters for the pulse are the same as in Fig.~\ref{fig:modes} and \ref{fig:transmission}; see also Section~\ref{sec:setup}.
	The SDR is 3.3$\mu m$ in radius and 180nm thick.
	Only for Fig.~\ref{fig:field}(b), a gold NP (20nm thick, 60nm radius) is placed on top of the SDR (at
	x=-2$\mu m$ and y=-$2\mu m$).
	This arrangement allows to excite a dipole plasmonic mode, illustrated as inset.}

Figure~\ref{fig:transmission} plots the pulse transmission for different arrangements of single NPs on top of an SDR.
For simplicity, and also to show-case the inherent potential for
highly nonlinear behavior of peo-PUFs, both simulations consider (a)~only one single gold NP versus no NP,\footnote{\blue{See
	Fig.~\ref{fig:transmission_5} for results concerning an arrangement of multiple NPs.}}
and (b)~no manufacturing variabilities, i.e., the SDR and the NP are assumed to be perfect disks without any roughness, etc.
Still, it is evident from
these simulations that already one perfect NP has a significant impact
on the propagation modes.
In reality, manufacturing variabilities and more NPs can
further enhance the randomness of peo-PUFs, thereby increasing their unpredictability.

Detrimental effects such as device aging, laser noise, or temperature variations may play a role in practice, but
this can only be properly investigated once peo-PUFs are manufactured, which is scope for future work.
In any case, error correction scheme to mitigate environmental impact on PUFs has been discussed in detail, e.g.,
   in \cite{maes15,herder14,ruehrmair13_IACR}.

\subsection{Key Generation and Authentication}
\label{sec:key_gen}

After applying a fixed input pulse, the output is processed as follows to obtain a key.
First, wave shaping/spectral filtering is applied to extract different partitions (or features) across the whole wavelength spectrum of the
optical response.
Second, analog-to-digital conversion is applied on each feature to obtain the underlying bits.
To enable a stable key, the noisy, least significant bits are rejected across all features.
Finally, for each feature, their stable bits are bundled, and these bundles are grouped.
Without loss of generality, the above steps are taken such that the key is 128 bits long.
In general, the number of key bits
dictates the number of bits to extract per feature and the number of features to be considered. This, in turn, dictates
the requirements for accuracy and resolution of the optoelectronics and the analog-to-digital circuitry.

For authentication of chips using integrated peo-PUFs,
we propose the following simple two-phase scheme, following the literature~\cite{maes10}.
During the initial \textit{enrollment phase}, which is to be conducted in a trusted environment,
the unique PUF responses are observed and different keys are derived.
Besides the actual keys, some key parameters---also known as \textit{helper data}---are to be recorded as well.
For peo-PUFs, that is: (a)~the selection of frequency features, and
(b)~the number of considered bits for each feature.
Note that this helper data is essential to verify the keys.
Furthermore, we record the \textit{fractional Hamming distance (FHD)} for the same key obtained under different operating conditions.
As for the actual \textit{authentication phase}, after enrollment, the PUFs are to be queried again,
using the same fixed input and a
selected configuration for the helper data, and the resulting key is compared with the recorded one.
The authentication is considered successful in case the key falls within the FHD expected for the operating conditions and helper data.

\section{Evaluation}
\label{sec:evaluation}

\subsection{Experimental Setup}
\label{sec:setup}

Simulations for the frequency and time domain are carried out using the \textit{Lumerical FDTD} software.
All simulation data is exported for the subsequent security evaluation.
If not specified otherwise, and without loss of generality,
we assume a 5$\mu m$ radius and 180nm height for the SDR,
a height of 30nm for the NPs,
an optical input pulse of 100fs duration,
an ambient temperature of 300K,
and a simulation time-frame of 8ps.
Note that each simulation took multiple days;
more NPs, longer simulation time-frames, etc., will significantly increase runtime.
For absorption and scattering phenomena (Fig.~\ref{fig:extinction}), \textit{COMSOL FEM} simulations were performed.

For the key-generation and authentication framework, we use \textit{Matlab} and custom scripts.
The security evaluation is based on \textit{entropy}, \textit{NIST randomness tests}~\cite{rukhin01}, and FHD~\cite{maes10}.
\blue{Without loss of generality, we leverage the frequency-domain data obtained from \textit{Lumerical FDTD} simulations for the security
	evaluation. We apply the post-processing steps outlined in Section~\ref{sec:key_gen}.
		We sample 1,000 different keys for each peo-PUF configuration under consideration, by varying the selection of frequency
		features and the number of bits per feature.
Note that we release our post-processing and security evaluation framework to the community via~\cite{webinterface}.}

We consider various peo-PUF setups, mainly for different arrangements of NPs, but also for different SDR tokens and operating
conditions (temperature and pulse width).  The configurations are labeled and summarized in Table~\ref{tab:configs}.

\begin{table}[tb]
\centering
\caption{
Considered peo-PUF configurations
}
\setlength{\tabcolsep}{0.30em}
\input{incl/tab-configs}
\label{tab:configs}
\end{table}

\subsection{Randomness}
\label{sec:randomness}

For peo-PUFs acting as weak PUFs, it is essential to quantify the underlying randomness, which also reflects on their
\textit{unpredictability}~\cite{maes10}.
As indicated, we evaluate the entropy and conduct NIST tests toward that end.

We report the entropy $S$ for the different peo-PUF
configurations in Table~\ref{tab:entropy-NIST}. All mean entropies are beyond 0.987, which hints on strongly random distributions
of zeroes and ones across all keys and for different peo-PUFs.
For the minimal entropies, however, we note that
\textit{Si5um} and \textit{Si5umTiN60nm}, i.e.,
the peo-PUF without any NP and the peo-PUF with one single titanium-nitride NP, respectively, appear the weakest.
Hence, the addition of plasmonic NPs can indeed help to increase randomness,
but the degree of randomness depends on the NP count, materials, etc.

\begin{table}[tb]
\centering
\caption{
Entropy and passing of NIST tests
}
\setlength{\tabcolsep}{0.40em}
\input{incl/tab-entropy-NIST}
\label{tab:entropy-NIST}
\end{table}

We furthermore conduct the following, commonly considered NIST tests~\cite{rukhin01}.
\begin{enumerate}
\item Frequency (F): testing for the proportion of zeros versus ones across the entire key.
\item Frequency within a block (FB): testing for the proportion of zeroes versus ones within $m$-bit blocks,
for $m=20$.
\item Runs (R): testing for the number of uninterrupted sequences (identical bits) across the entire key.
\item Approximate entropy (AE): testing for the frequency of all possible $m$-bit patterns, with respect to
an enumeration of all possible overlapping blocks of consecutive lengths ($m$ and $m+1$), for $m=3$.
\end{enumerate}
The NIST tests are all based on \textit{p-values},
i.e., they can quantify the confidence for passing (or failing) particular test.
More specifically, the p-values represent the probability that a perfect random number generator would have produced a
sequence less random than the sequence that was tested.
In other words, if the p-value for a particular test is equal to 1, then the key appears to be perfectly
random concerning the kind of randomness assessed by that test.

In Table~\ref{tab:entropy-NIST}, we report the percentage of keys passing the NIST tests, all for a confidence interval of 99\%. 
We observe that the configurations \textit{Si5umAu60nm}, \textit{Si5um}, \textit{Si5umAu60nm(c)}, and \textit{Si5umTiN60nm} are
inferior to others.
Hence, configurations with none or only one NP are limited.
Here we would like to caution that already the positioning of single NPs may lead to these results.
That is corroborated by the fact that the configuration \textit{Si5umAu60nm(b)} is superior, although it also holds only one gold NP
of same size and shape as the other configurations.
In reality, where we are free from considerable simulation runtimes,
peo-PUFs could and should comprise many more NPs,
all randomly arranged.
We believe that the arrangement of an individual NP would play no significant, possibly deteriorating role there anymore for
the overall randomness.

\subsection{Uniqueness and Reproducibility}
\label{sec:uniqueness}

Two further key properties for any PUF are \textit{uniqueness}
and \textit{reproducibility}~\cite{maes10}.
Both uniqueness and reproducibility are to be measured on pairs of PUF outputs resulting from the same challenge.
Uniqueness describes the difference of outputs across two PUF instances,
whereas reproducibility describes the similarity of outputs for the same PUF instance, but under different 
operating conditions.
Therefore, reproducibility can also be thought of as reliability.
The FHD, short for
fractional Hamming distance,
	is used to quantify both properties.

Regarding FHD for uniqueness, also known as \textit{inter-FHD}, the ideal value is 50\%;
regarding FHD for reproducibility, also known as \textit{intra-FHD}, the ideal value is 0\%.
Since the inter- and intra-FHD can vary depending on the applied challenge/helper data,
we report Gaussian FHD distributions along with their histograms, mean values $\mu$, and standard deviations $\sigma$
(as suggested in~\cite{maes10}).
It is important to note that deviations from ideal inter-/intra-FHD values are tolerable as long as their distributions remain
reasonably separated.

Next, we investigate peo-PUFs for
two critical operation parameters, namely input pulse width and ambient
temperature.

\begin{figure}[tb]
\centering
\includegraphics[width=.85\columnwidth]{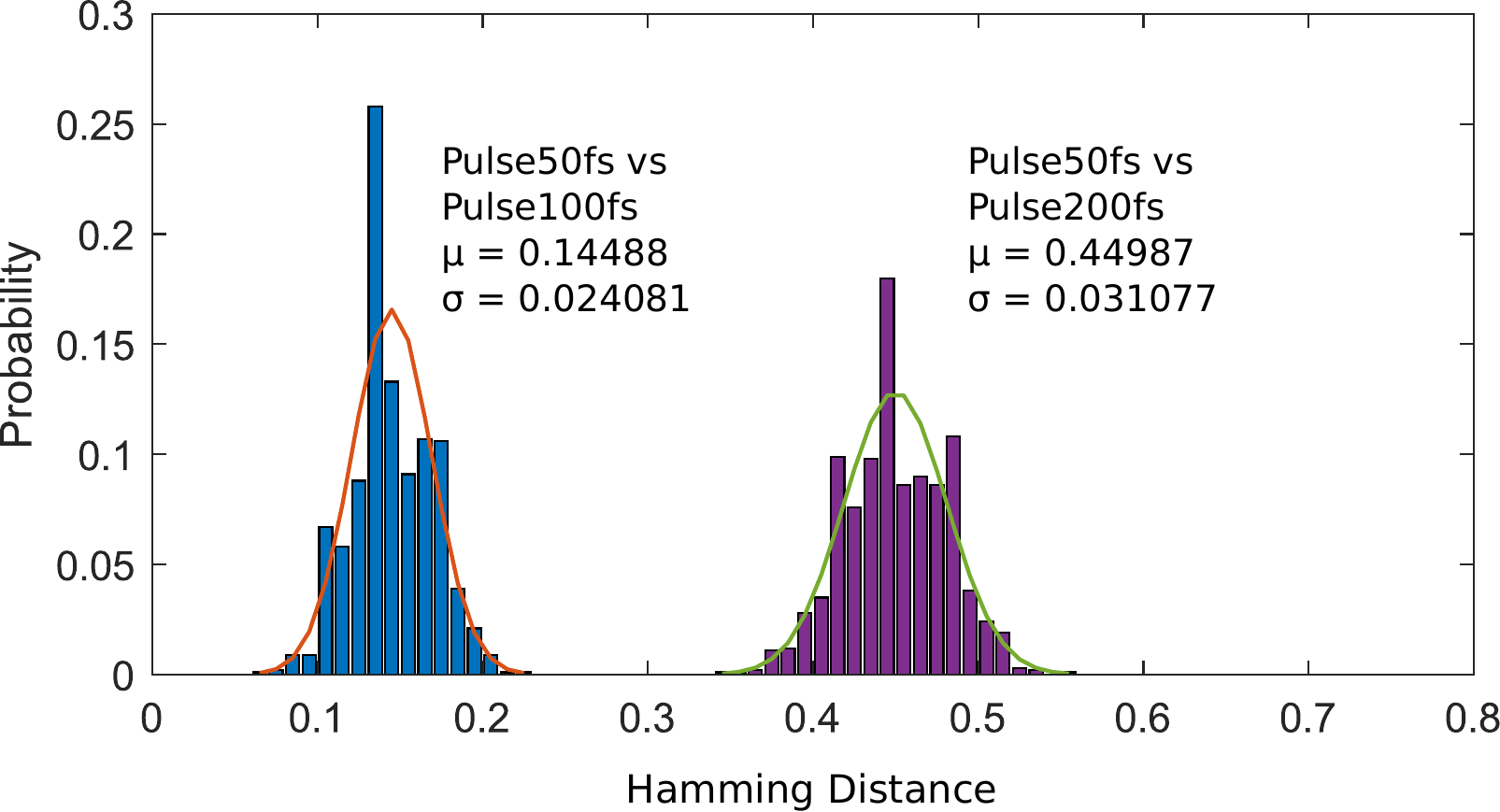}
\caption{Intra- and inter-FHD for different input pulses.
\label{fig:FHD_pulses}
}
\end{figure}

We contrast the FHD distribution for different pulse widths
in Fig.~\ref{fig:FHD_pulses}.
That is, here we assume that the input may exhibit some noise which peo-PUFs should be able to tolerate,
at least to some degree.
More specifically,
     we consider the scenario \textit{Pulse100fs} versus \textit{Pulse50fs}
as tolerable fluctuations for one and the same input challenge, i.e., concerning reproducibility and intra-FHD.
Another scenario,
\textit{Pulse200fs} versus \textit{Pulse50fs}, is considered as
comparing two different peo-PUFs with different laser setups, i.e., concerning uniqueness and inter-FHD.
Now, from Fig.~\ref{fig:FHD_pulses}, we note that the FHD distributions for these two scenarios are clearly distinct. Hence, the peo-PUFs
are (a) reproducible for small input variations, around 50fs, and (b) unique for different laser setups.

\begin{figure}[tb]
\centering
\includegraphics[width=.85\columnwidth]{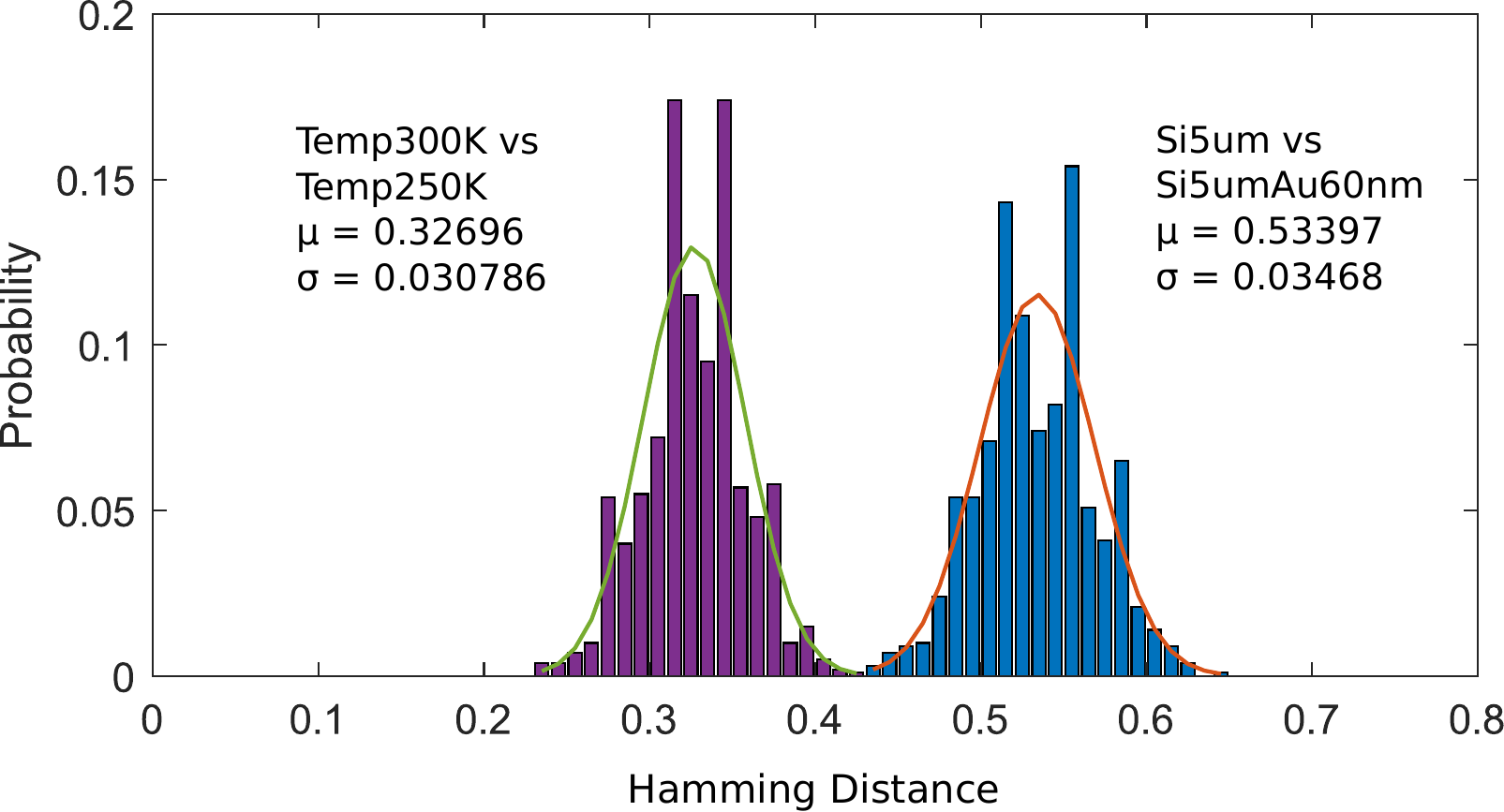}
\caption{Intra-FHD for different ambient temperatures but for the same peo-PUF, versus inter-FHD for different peo-PUFs.
	For the latter, note that the underlying spectra are provided in Fig.~\ref{fig:transmission}.
\label{fig:FHD_temp}
}
\end{figure}

The ambient temperature impacts the reproducibility of most, if not all, types of PUFs~\cite{maes10,chang17_PUF,herder14}.
In Fig.~\ref{fig:FHD_temp}, we contrast the intra-FHD for the same peo-PUF (with one gold NP) at 300K versus 350K ambient
temperature, after applying correlation-based shifting of the wavelength spectra.
While we observe more noise than it was the case for the reproducibility under input-pulse
fluctuations, the intra-FHD distribution still remains separated from another inter-FHD distribution (\textit{Si5umAu60nm} versus
\textit{Si5um}) which was obtained from different peo-PUFs.
Hence, peo-PUFs can tolerate some temperature fluctuations, although further compensation measures may be required in practice,
where other noises such as voltage glitches may play some role as well.

\begin{figure}[tb]
\centering
\includegraphics[width=.85\columnwidth]{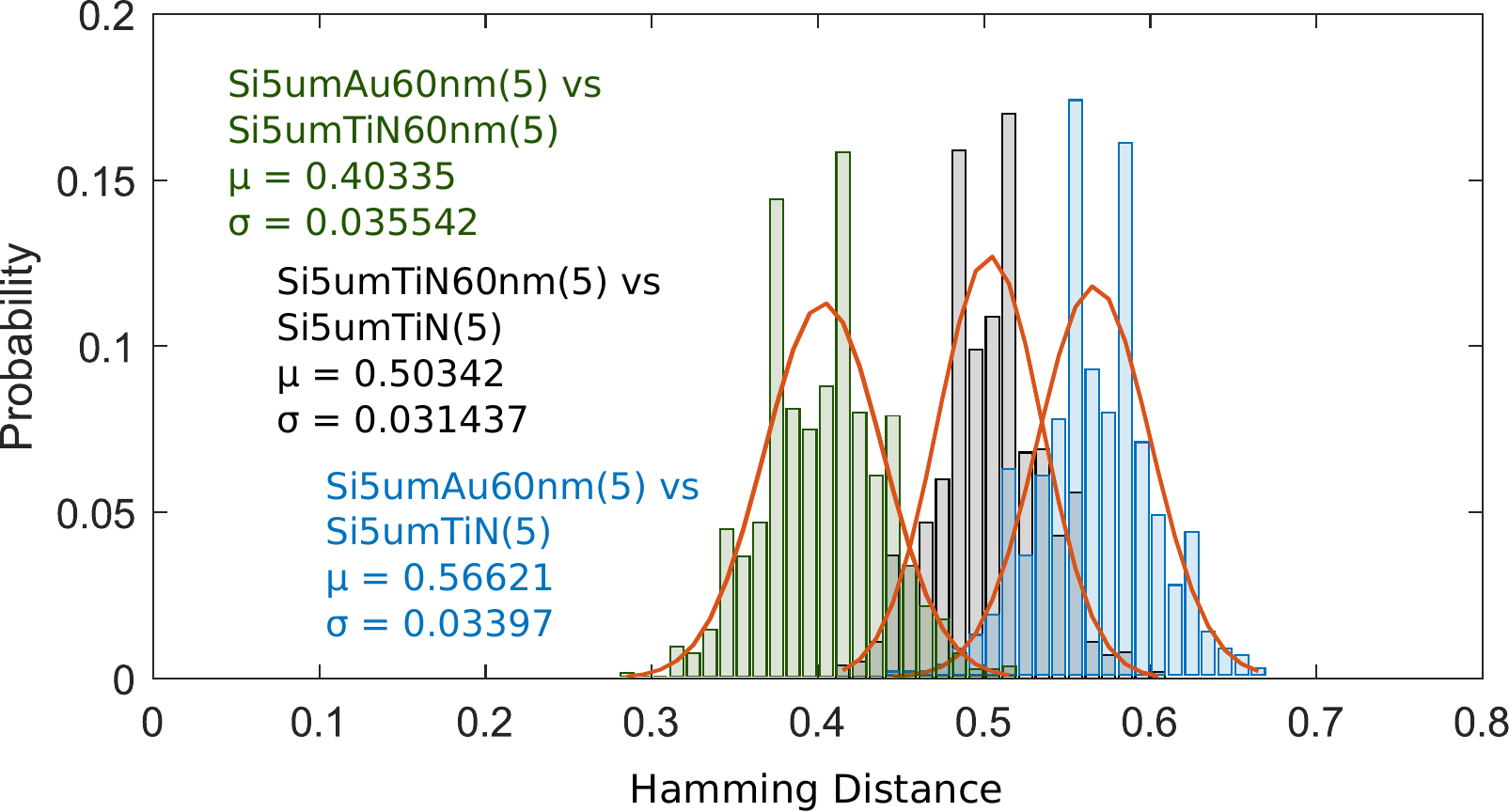}
\caption{Inter-FHD for five NPs.
	\blue{The underlying spectra are provided in Fig.~\ref{fig:transmission_5}.}
\label{fig:FHD_5_NPs}
}
\end{figure}

\begin{figure}[tb]
\centering
\sidesubfloat[]{
	\includegraphics[height=42mm, trim = {9mm 0mm 5mm 1mm}, clip=true]{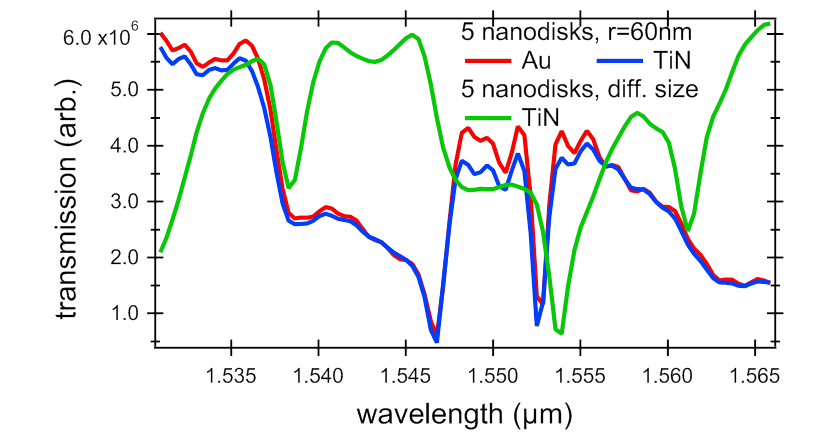}
}
\\[2mm]
\sidesubfloat[]{
	\includegraphics[height=42mm, trim = {8mm 0mm 5mm 1mm}, clip=true]{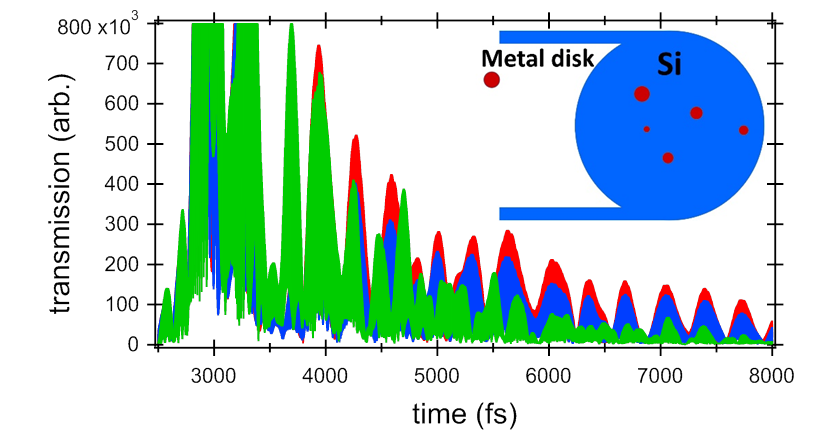}
}
\caption{\blue{Transmission plots for different peo-PUF tokens with five NPs.
	The red, blue, and green curves correspond to the setups \textit{Si5umAu60nm(5)}, \textit{Si5umTiN60nm(5)},
	    and \textit{Si5umTiN(5)}, respectively, as described in Table~\ref{tab:configs}.
		The inset in (b) shows the spatial arrangement of NPs for \textit{Si5umTiN60nm(5)}, with NPs labeled as ``metal disk.''}
\label{fig:transmission_5}
}
\end{figure}

Regarding uniqueness, besides the configurations already covered in Fig.~\ref{fig:FHD_pulses} and~\ref{fig:FHD_temp},
we investigated further peo-PUFs.
In Fig.~\ref{fig:FHD_5_NPs}, we provide three inter-FHD distributions for exemplary arrangements of five NPs.
   The means range from 0.4 to 0.56, with reasonably low standard deviations of 0.03.
The distributions attest to the potential for strong uniqueness of peo-PUFs.
It should be emphasized again that, in reality, considerably more than five NPs will be present.
Therefore, the inter-FHD distributions and uniqueness can be expected to improve even further.
\blue{The transmission plots as well as one arrangement of NPs related to Fig.~\ref{fig:FHD_5_NPs} are illustrated in
Fig.~\ref{fig:transmission_5}.  Note that most of the NPs were placed in the middle of the SDR (inset Fig.~\ref{fig:transmission_5}(b)),
where the interaction of the propagating photonic mode with NPs is relatively weak.  This interaction can be
largely enhanced through simple manufacturing means, e.g., by placing a metallic scatterer inside the SDR, to raise the uniqueness of peo-PUFs even further.}

\section{Conclusion}

Ensuring the reliability and trustworthiness of 
electronic systems has become a pressing concern nowadays.
The notion of physically unclonable functions (PUFs) can serve
to implement challenge-response-based authentication schemes or 
``device fingerprinting,'' i.e., device-specific key generation.
While most PUFs have eventually fallen short, especially in terms of unpredictability,
optical PUFs are widely considered as promising; such 
PUFs derive their randomness from optical phenomena, independent of the underlying microelectronics.

In this work, for the first time, we explore the use of plasmonic NPs toward enhanced optical PUFs, called peo-PUFs.
In general, PUFs have to be unpredictable, unique, and unclonable, yet reproducible.
The core principle of peo-PUFs---the random depositioning of various NPs on top of a nonlinear silicon disc resonator---can serve
well to achieve all these requirements.
That is because the intertwining of silicon photonics and plasmonics by construction enables highly nonlinear behavior, which is
unpredictable, unique, and overly difficult to clone/reproduce.
We have explored that in a first-of-its-kind, physics-simulation-based study for different peo-PUF configurations and conditions, thereby
also confirming reproducibility.
We have conducted a security analysis of peo-PUFs in the context of key generation and authentication.

For future work, we plan to manufacture and
characterize peo-PUFs.
Toward that end, we also call for relatively simple optoelectronics, which may allow for fully monolithic integration of peo-PUFs on silicon
chips.
Overall, we believe that the vast range of readily available NPs can open up a unique opportunity for constructing highly
resilient PUFs.

\input{main.bbl}

\begin{IEEEbiographynophoto}{Johann Knechtel}
received the M.Sc.\ in Information Systems Engineering (Dipl.-Ing.) in 2010 and the Ph.D.\ in Computer Engineering
(Dr.-Ing.) in 2014, both from TU Dresden, Germany.  He is currently a Research Associate
at the New York University Abu Dhabi (NYUAD), UAE.  Dr.\ Knechtel was a
Postdoctoral Researcher in 2015--16 at the Masdar Institute of Science and Technology, Abu Dhabi.  From 2010 to 2014, he was
a Scholar with the DFG Graduate School on ``Nano- and Biotechnologies for Packaging of Electronic
Systems'' and the Institute of Electromechanical and Electronic Design, both hosted at the TU Dresden.  In 2012, he was a
Research Assistant with the Dept.\ of Computer Science and Engineering, Chinese University of Hong Kong, China.  In 2010, he
was a Visiting Research Student with the Dept.\ of Electrical Engineering and Computer Science, University of Michigan, USA.
His research interests cover VLSI Physical Design Automation, with particular focus on Emerging Technologies and Hardware Security.
\end{IEEEbiographynophoto}

\begin{IEEEbiographynophoto}{Jacek Gosciniak}
received the M.Sc.\ in Applied Physics in 2002 from Technical University of Lodz, Poland, and the Ph.D.\ in Functional Materials and Nanotechnology
in 2012 from University of Southern Denmark, Denmark. He is currently a Research Associate at the New York University Abu Dhabi (NYUAD), UAE. He
has industry experience at Spectra-Physics, Newport Corp., USA where he spend over one and an half year working as the Application Lab Engineer.
Dr.\ Gosciniak was a Postdoctoral Researcher in 2012--13 at the Singapore University of Technology and Design, Singapore, where he was involved in
graphene-based plasmonic modulators.

Dr.\ Gosciniak has extensive experience in working in national and European research projects on plasmonic components and devices, integration of
plasmonics with photonics platform and implementation of plasmonics into data storage devices like PLASMOCOM, PLATON, COMPASS and ANAP (national
		project). Furthermore, his research potential has been recognized by the EU as he was awarded a Marie Curie Research Fellowship
at Tyndall National Institute, where he has succeeded in bringing together a team consisting of both industrial (Seagate Technology) and academic
partners (Tyndall National Institute). 
\end{IEEEbiographynophoto}

\begin{IEEEbiographynophoto}{Alabi Bojesomo}
obtained the M.Sc.\ in Microsystems Engineering from Masdar Institute of Science and Technology Abu Dhabi in 2016, and a B.Sc.\ in
Electrical Engineering from Obafemi Awolowo University (OAU) Ile-Ife in 2011.  He is currently a Research Engineer at the New York University Abu
Dhabi (NYUAD), UAE. His research interests include MEMS and hardware security.
\end{IEEEbiographynophoto}

\begin{IEEEbiographynophoto}{Satwik Patnaik}
received B.E.\
in Electronics and Telecommunications from the University of
Pune, Pune, India and
M.Tech.\ in Computer Science and Engineering with a specialization in VLSI Design from Indian Institute of Information Technology and
Management, Gwalior, India. 
He is a Ph.D.\ candidate at the Department of Electrical and Computer Engineering at the 
Tandon School of Engineering with New York University, Brooklyn, 
NY, USA. 
He is a Global Ph.D.\ Fellow with New York University Abu Dhabi, Abu Dhabi, UAE. 
His current research interests 
include Hardware Security, Trust and Reliability issues for CMOS and Emerging Devices 
with particular focus on low-power VLSI Design.
He is a student member of IEEE and ACM.
\end{IEEEbiographynophoto}

\begin{IEEEbiographynophoto}{Ozgur Sinanoglu}
is a Professor of Electrical and Computer Engineering at New York University Abu Dhabi. He earned his B.S.\ degrees, one in
Electrical and Electronics Engineering and one in Computer Engineering, both from Bogazici University, Turkey in 1999. He obtained his MS
and PhD in Computer Science and Engineering from University of California San Diego in 2001 and 2004, respectively. He has industry
experience at TI, IBM and Qualcomm, and has been with NYU Abu Dhabi since 2010. During his PhD, he won the IBM PhD fellowship award twice.
He is also the recipient of the best paper awards at IEEE VLSI Test Symposium 2011 and ACM Conference on Computer and Communication Security
2013. 

Prof.\ Sinanoglu's research interests include design-for-test, design-for-security and design-for-trust for VLSI circuits, where he has more
than 180 conference and journal papers, and 20 issued and pending US Patents. Sinanoglu has given more than a dozen tutorials on hardware
security and trust in leading CAD and test conferences, such as DAC, DATE, ITC, VTS, ETS, ICCD, ISQED, etc. He is serving as track/topic
chair or technical program committee member in about 15 conferences, and as (guest) associate editor for IEEE TIFS, IEEE TCAD, ACM JETC,
      IEEE TETC, Elsevier MEJ, JETTA, and IET CDT journals. 

Prof.\ Sinanoglu is the director of the Design-for-Excellence Lab at NYU Abu Dhabi. His recent research in hardware security and trust
is being funded by US National Science Foundation, US Department of Defense, Semiconductor Research Corporation, Intel Corp and
Mubadala Technology.
\end{IEEEbiographynophoto}

\begin{IEEEbiographynophoto}{Mahmoud Rasras}
is an Associate Professor of the Electrical and Computer Engineering at New York University Abu Dhabi (NYUAD). He received a PhD degree in
physics from the Catholic University of Leuven, Belgium.  Dr.\ Rasras has more than 11 years of industrial research experience as a Member of
Technical Staff at Bell Labs, Alcatel-Lucent, NJ, USA. Prior to joining NYUAD. Dr.\ Rasras was a faculty member and former Director of the SRC/GF
Center-for-Excellence for Integrated Photonics at Masdar Institute (part of Khalifa University). He authored and co-authored more than 120
journal and conference papers and holds 33 US patents. Dr.\ Rasras is an Associate Editor of Optics Express, Guest Editor -- MDPI, and a Senior
IEEE Member.
\end{IEEEbiographynophoto}

\end{document}

%% file: incl/tab-configs.tex
\begin{tabular}{l||l}
Label & Short Description \\ \hline
1) Si5umAu60nm(a)     & SDR radius 5$\mu m$, 1 gold NP radius 60nm   \\
2) Si5umAu60nm(b)    & As Si5umAu60nm(a), but different location for NP \\
3) Si5umAu60nm(c)   & As Si5umAu60nm(a), but different location for NP \\
4) Si5umAu60nm(5)    & As Si5umAu60nm(a), but 5 gold NP \\
5) Si6umAu60nm     & SDR radius 6$\mu m$, 1 gold NP radius 60nm  \\
6) Si5umAu120nm    & SDR radius 5$\mu m$, 1 gold NP radius 120nm  \\
7) Si6umAu120nm    & SDR radius 6$\mu m$, 1 gold NP radius 120nm  \\
8) Si5umAu240nm    & SDR radius 5$\mu m$, 1 gold NP radius 240nm  \\
9) Si6umAu240nm    & SDR radius 6$\mu m$, 1 gold NP radius 240nm  \\
10) Si5um		& SDR radius 5$\mu m$, no NP \\
11) Pulse100fs 	& As Si5umAu60nm(a), but different location for NP  \\
12) Pulse50fs  	& As Pulse100fs, but input pulse width 50fs \\
13) Pulse200fs 	& As Pulse100fs, but input pulse width 200fs \\
14) Temp300K 	& As Si5umAu60nm(a), but different location for NP	  \\
15) Temp350K 	& As Temp300K, but ambient temperature 350K	  \\
16) Si5umTiN60nm	& As Si5umAu60nm(a), but 1 titanium nitride NP \\
17) Si5umTiN60nm(5)	& As Si5umAu60nm(a), but 5 titanium nitride NP \\
18) Si5umTiN(5)	& As Si5umAu60nm(a), but 5 titanium nitride NP \\
		& with radius ranging from 60nm to 240nm \\
\end{tabular}

%% file: incl/tab-entropy-NIST.tex
\begin{tabular}{l||l|l||l|l|l|l}
& Min $S$ & Mean $S$ & F (\% )& FB (\%) & R (\%) & AE (\%) \\ \hline
Si5umAu60nm(a)     & 0.933   & 0.9911 & 83.8      & 98.1           & 99.8 & 98.9               \\
Si5umAu60nm(b)    & 0.9625  & 0.9972 & 98.6      & 100            & 83.9 & 96.5               \\
Si5umAu60nm(c)   & 0.928   & 0.9871 & 72.3      & 99.1           & 99.9 & 98.1               \\
Si5umAu60nm(5)    & 0.9575  & 0.9968 & 98.2      & 99.4           & 99.8 & 99.3               \\
Si6umAu60nm     & 0.9727  & 0.997  & 98.5      & 99.9           & 98.3 & 98.8               \\
Si5umAu120nm    & 0.9579  & 0.9969 & 98.4      & 100            & 93.1 & 98                 \\
Si6umAu120nm    & 0.9597  & 0.9979 & 98.6      & 100            & 98.7 & 98.7               \\
Si5umAu240nm    & 0.9682  & 0.9961 & 96.3      & 99.2           & 99.1 & 98.4               \\
Si6umAu240nm    & 0.9523  & 0.9963 & 97        & 99.6           & 100  & 99.8               \\
Si5um		& 0.9259  & 0.9926 & 88.1      & 99.4           & 99.9 & 98.5               \\
Pulse100fs 	& 0.964   & 0.9964 & 97.3      & 100            & 90.9 & 97.4               \\
Pulse50fs  	& 0.9681  & 0.9978 & 98.6      & 100            & 83.8 & 96.5               \\
Pulse200fs 	& 0.9696  & 0.9965 & 98.8      & 99.7           & 100  & 99.3               \\
Temp300K 	& 0.9632  & 0.996  & 97        & 99.8           & 97.7 & 97.5               \\
Temp350K 	& 0.9258  & 0.9905 & 81.7      & 99.7           & 100  & 98.8               \\
Si5umTiN60nm	& 0.9183  & 0.9895 & 80.6      & 98.7           & 99.7 & 97.9               \\ 
Si5umTiN60nm(5)	& 0.9398  & 0.9902 & 83.6      & 99.2           & 99.4 & 98.2               \\
Si5umTiN(5)	& 0.9423  & 0.9911 & 89.9      & 98.8           & 99.8 & 98.5               \\ 
\end{tabular}